\documentclass[fleqn,twoside]{article}
\usepackage[headings]{espcrc2}

\readRCS
$Id: espcrc2.tex,v 1.2 2004/02/24 11:22:11 spepping Exp $
\usepackage{graphicx}
\usepackage[figuresright]{rotating}

\newcommand{\AmS}{{\protect\the\textfont2
  A\kern-.1667em\lower.5ex\hbox{M}\kern-.125emS}}

\def\GeV{\rm GeV}

\def\lapproxeq{\lower .7ex\hbox{$\;\stackrel{\textstyle
<}{\sim}\;$}}
\def\gapproxeq{\lower .7ex\hbox{$\;\stackrel{\textstyle
>}{\sim}\;$}}

\title{Summary of the HERA-LHC workshop}

\author{R. S. Thorne\address[MCSD]{Department of Physics and Astronomy, 
        University College London, Gower Street, 
        London, WC1E 6BT, United Kingdom}}%
       
\begin{document}
\begin{abstract}
\vspace{1pc}
I present a summary of the last in the series of HERA-LHC workshops, 
CERN, 26-30th May 2008. 
\end{abstract}

\maketitle

\section{Introduction}

In May 2008 the final stage of a 4-year series of meetings at DESY and 
CERN took place. The full timetable took the form
\begin{itemize}

\item 26-27 March 2005 CERN (250-300 participants)

\item 6-9 June 2006 CERN (150 participants)

\item 12-16 March 2007 DESY (160 participants) 

\item 26-30 May 2008 CERN (190 participants)

\end{itemize}
with a variety of additional smaller meetings,  e.g. working weeks and 
working  group meetings. The workshop was organised into 5 core working 
groups. 

\begin{itemize}

\item Parton density functions

\item Multi-jet final states and energy flows

\item Heavy quarks (charm and beauty)

\item Diffraction

\item Monte Carlo tools. 

\end{itemize}

I have been rather more involved with some of these groups than others, 
and am more qualified to talk on some than others. 
This summary will obviously reflect this, and will largely be highlights 
rather than a list. 
I will present the summary roughly by topic, but not always exactly 
by the Working Group, and indeed, the many overlapping sessions makes this 
impossible. I will also try not to dwell on topics covered in detail in 
other contributions to these proceedings.

\section{Parton distributions}

\begin{figure}[htb]
\vspace{9pt}
  \includegraphics[width=0.48\textwidth]{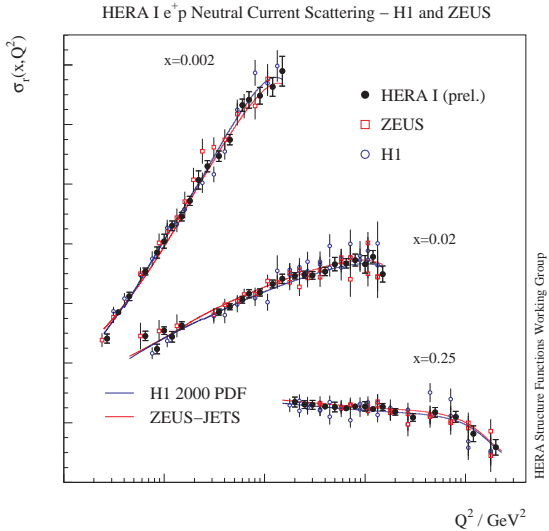}
\caption{The averaged HERA structure function data in a sample of $x$ bins.}
\label{fig:avHERA}
\end{figure}

One of the main foci of the workshop was the issue of the extraction of the 
parton distributions, largely by using HERA data, and the consequences for LHC
predictions. At this final meeting a number of new data sets were presented.
The most interesting ``new'' data were clearly those on $F_L(x,Q^2)$, 
where it was shown that the earliest data \cite{Aaron:2008tx} are consistent 
with NLO and NNLO predictions, but where the lower $Q^2$ data yet to be 
analysed could show some sensitivity to extended theoretical approaches. 
However, potentially the most important data presented 
(probably at the whole meeting) were the averaged HERA measurement of 
the total inclusive cross section \cite{averagedata}. An example is shown in 
Fig. \ref{fig:avHERA}. As well as an improvement in statistics  there is 
a significant elimination of correlated errors and the data has an
accuracy of $1\,-\,2\%$ over an enormous range of both $x$ and $Q^2$. 
When finally published it will be one of the best tests of factorization 
and perturbative QCD that we have. There is already a fit to these data 
performed by the experiments, with an impressive reduction in 
uncertainties \cite{Reisert:2008ip}. 
However, in comparing with other distributions there 
are lots of things to consider, e.g. different heavy flavour treatments, 
and particularly the number of free parameters used, which is small
in these fits and likely to lead to constraints.

\begin{table*}
  \caption{Ratios of predictions for $W\equiv W^++W^-$ and $Z$ total cross 
sections at the Tevatron ($\sqrt{s} = 1.96$ TeV) and LHC ($\sqrt{s} = 14$ TeV)
calculated using the central values of different PDF sets, with respect to 
those from the MSTW 2008 sets.}
  \label{tab:wztotpdf}
  \newcommand{\m}{\hphantom{$-$}}
\newcommand{\cc}[1]{\multicolumn{1}{c}{#1}}
\renewcommand{\tabcolsep}{2pc} 
\renewcommand{\arraystretch}{1.2} 
\begin{tabular}{@{}l|cc|cc}
    \hline
    & \multicolumn{2}{c|}{Tevatron} & \multicolumn{2}{c}{LHC} \\
    Ratio to MSTW 2008 & $\sigma_W$ & $\sigma_Z$ & $\sigma_W$ & $\sigma_Z$ \\
    \hline
    MRST 2006 NNLO              & 1.00 & 1.01 & 0.99 & 1.00 \\ 
    MRST 2004 NNLO              & 0.99 & 1.00 & 0.93 & 0.94 \\ \hline
    MRST 2006 NLO (unpublished) & 0.99 & 1.00 & 1.00 & 1.01 \\
    MRST 2004 NLO               & 0.99 & 1.00 & 0.97 & 0.98 \\
    CTEQ6.6 NLO                 & 0.98 & 0.99 & 1.02 & 1.02 \\
    \hline
\end{tabular}\\[2pt]
\end{table*}

\begin{figure}[htb]
\vspace{9pt}
  \includegraphics[width=0.48\textwidth]{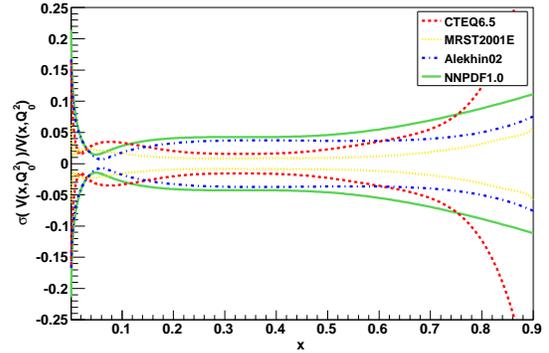}
\caption{The NNPDF1.0 valence quark uncertainty compared to other sets.}
\label{fig:NNPDF}
\end{figure}

While waiting for these averaged data the PDF-fitting groups provided 
presentations outlining many improvements and updates. 
There was a detailed 
discussion of the CTEQ6.6 PDFs \cite{cteq66}, which contain a number of
major changes. In particular it adopts the general-mass heavy flavour 
scheme as default (as does CTEQ6.5 \cite{cteq65}) 
and fits directly  to strange quarks. There are now $44$ (previously $40$) 
eigenvector sets for the PDFs with uncertainties. The MSTW collaboration 
also presented a preliminary 2008 set \cite{MSTW08} based on a fit to a very 
wide variety of new data. They now also fit  the strange quark distribution 
directly but in this case it is much closer to the  
previous distribution, though the extra free parameters increase the 
uncertainties on all the light sea quarks. They now produce $40$ (previously 
$30$) eigenvector sets. The main MSTW focus was on 
a new {\it dynamical} determination of the tolerance, i.e. the  
$\Delta \chi^2$ used to determine the uncertainty.
There is no longer a fixed value,  but a systematic 
analysis eigenvector by eigenvector.
They also highlighted the wide variety of new 
data included in the fit. Most important is the inclusion of 
new Tevatron data.
This gives detailed information on quark decomposition, since it 
probes different weightings from the structure functions --
for example, $d_V(x,Q^2)$ is now a different type of shape.
Additionally the CDF and D0 Run II inclusive jet data in 
different rapidity bins \cite{Abulencia:2007ez,Abazov:2008hu,Aaltonen:2008eq}
is fit very well and MSTW find a 
much softer high-$x$ gluon with these new data. One of the main consequences 
of these improvements in both analyses is a convergence of the predictions for 
cross sections at the LHC, as shown in Table~\ref{tab:wztotpdf}.

\begin{table*}
  \caption{A summary of the potential for luminosity measurements at LHCb 
using various measurements and techniques.}
  \label{tab:lumi}
  \newcommand{\m}{\hphantom{$-$}}
\renewcommand{\arraystretch}{1.2} 
\begin{tabular}{@{}l|c|c|c}
    \hline
    & 2008(5pb$^{-1}$) & 2009(0.5fb$^{-1}$)& 2010(2fb$^{-1}$)\\
    \hline
    Van Der Meer             & 20$\%$ & 5-10$\%$ & 5-10$\%$ \\ 
    Beam Gas                 & 10$\%$ & 5$<\%$ & 5$<\%$ \\ 
    $Z \to \mu^+\mu^-$       & 5$\%$ & 4$\%$ & 4$\%$ \\ 
    $pp \to pp + \mu^++\mu^-$  & 20$\%$ & 2.5$\%$ & 1.5$\%$ \\ 

    \hline
\end{tabular}\\[2pt]
\end{table*}

There was also the presentation of an entirely new set of PDFs by the 
NNPDF group \cite{NNPDF}, obtained by fitting to DIS data. 
This relaxes the restriction in uncertainty due to fixed 
parameterizations, though in practice relies on a very large 
number of parameters. It also proceeds by making
a Monte Carlo sample of the distribution of experimental data by generating 
a large number of replicas of data centred on each data point with full 
inclusion of the information from errors and their correlations. Each replica 
is used to generate a PDF set, and the mean for a quantity is obtained by 
averaging, and the uncertainties from standard deviations. 
The data are split into training and validation sets for each 
replica, and the $\chi^2$ for one monitored while the other is minimised. 
This avoids over-complicating the input PDFs by stopping when the fit to 
the validation sets stops improving. The final fit quality 
seems slightly worse than in the global fits. 
The PDFs obtained seem to be fairly similar to the existing global fit sets 
(note at present they use a ZM-VFNS and more limited quark decomposition).
The uncertainties for valence quarks are shown in Fig. \ref{fig:NNPDF}.
In the region where the data are most constraining the uncertainty is quite 
similar to other approaches, but does tend to expand beyond this 
in regions where extrapolation is required, e.g. very high and low $x$. 
It must be remembered that there is less constraint from data than for CTEQ
and MSTW at present. It will be interesting to see future developments.  

\section{Standard candles}

The workshop included a special section on this topic. One of the central 
features was a presentation by representatives of each of ATLAS, CMS and LHCb 
on luminosity measurements. A detailed breakdown of expectations for LHCb
can be seen in Table~\ref{tab:lumi} \cite{lumitable}. 
To summarise, at LHCb, ATLAS and CMS, in 
very early running total $Z$ boson production is probably the best 
determination since the theoretical uncertainty on the cross section 
including all uncertainty sources is $4-5\%$. Later a measurement of
$pp \to pp + \mu^++\mu^-$ may provide the best long-term determination at 
LHCb, possibly $\sim 2\%$, but anything comparable at ATLAS and CMS will 
require total and elastic cross section measurements using forward detectors,
e.g. ALFA (ATLAS) and TOTEM (CMS), and dedicated short low luminosity runs 
using particular beam optics. This could achieve a luminosity determination 
of accuracy $3-5\%$ after one year.

\begin{figure}[htb]
  \includegraphics[width=0.48\textwidth]{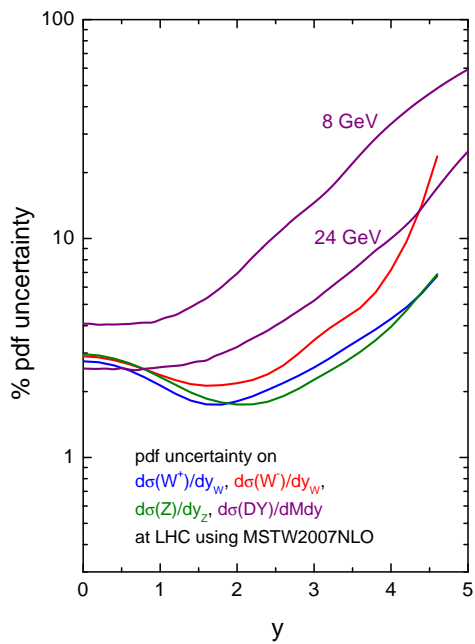}
\caption{The uncertainty from PDFS on vector boson production.}
\label{fig:WZrap}
\end{figure}

It was also suggested by the CTEQ collaboration \cite{cteq66} that the 
total $t \bar t$ cross section could be a useful standard candle, 
with both theory
and data uncertainties approaching $5\%$. However, a presentation by Mangano
\cite{top} was in contradiction to this. The most-up-to-date predictions
using NLO plus next-to-leading threshold logarithms 
for the $t \bar t$ cross section at $14$ TeV at the LHC for 
$m_t = 171~\mathrm{GeV}$ were shown to be
\[
\sigma_{t\bar t}
= 908 {~}^{+82(9.0\%)}_{-85(9.3\%)}~\mathrm{(scales)} 
{~}^{+30(3.3\%)}_{-29(3.2\%)} ~\mathrm{(PDFs)} ~~\mathrm{pb} 
\]
using CTEQ6.5 PDFs \cite{cteq65} and 
\[
\sigma_{t\bar t}
= 961 {~}^{+89(9.2\%)}_{-91(9.4\%)}~\mathrm{(scales)} 
{~}^{+11(1.1\%)}_{-12(1.2\%)} ~\mathrm{(PDFs)} ~~\mathrm{pb}
\nonumber
\]
using MRST2006 PDFs \cite{MSTW}.
Hence, the scale uncertainty of order $9\%$. Addtionally, although the 
PDF uncertainty using each set is much smaller, the central values differ
by $\sim 6\%$, due to systematic differences in 
the groups parameterisations and heavy flavour schemes. The effect from the 
latter may well be smaller with updated sets. Doubt was also expressed about
the 
accuracy of the measurement reaching $5\%$ for some time, but in the long term
$t\bar t$ production may help constrain the gluon for $0.01<x<0.1$.  

In the short term the best constraint on PDFs and the most precise comparisons
of data and theory will be for electroweak boson production. The 
uncertainty on various cross sections due to PDFs is shown as a function of 
rapidity in Fig. \ref{fig:WZrap}. The uncertainty on $\sigma(Z)$ and 
$\sigma(W^+)$ grows at high rapidity, that on $\sigma(W^-)$ grows more 
quickly at very high $y$ since it depends on the less well-known down quark,
and the uncertainty on $\sigma(\gamma^{\star})$ is greatest as 
$y$ increases, since it depends on the poorly know (even after HERA) 
partons at very small $x<10^{-4}$ \cite{LHCbpred}. 
Measurements of all these bosons (and their ratios) will be good enough in 
one detector or another to put new constraints on PDFs quite quickly. 
This has the potential to be most dramatic for high rapidity, 
low mass Drell-Yan production which is possible at LHCb \cite{LHCbgam}. 
However, it was highlighted at the meeting that fully reliable calculations of 
all of these processes require great care. For real precision account
of electroweak corrections must be made \cite{EW}. Also, small-$x$ 
resummations could give effects of a few $\%$, or possibly more for 
low mass Drell Yan. There were presentations from the groups working in this 
field \cite{sx1,sx2,sx3}. Though the groups are at various stages of 
development, it seems clear there is broad agreement on qualitative 
features, but some variation in approaches which lead to differences in 
details of results. It was pointed out that the measurements of $F_L(x,Q^2)$
at smaller $Q^2$ than has yet been analysed could give some information on
resummation effects compared to fixed order calculations. They could also 
provide information on the success of dipole models. 

\section{Dipole models, saturation}

\begin{figure}[htb]
\vspace{9pt}
  \includegraphics[width=0.48\textwidth]{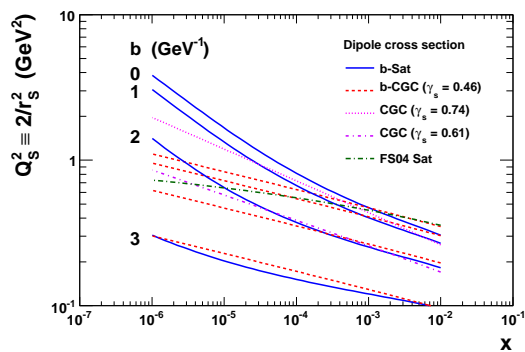}
\caption{The saturation scale for a variety of dipole models.}
\label{fig:satscale}
\end{figure}

Dipole models overlap with small-$x$ resummation, including some of 
the same corrections beyond a fixed order calculation. They are based on a less
complete framework than the collinear factorization approach, but can be used 
in a wider region since they apply at low $Q^2$. A certain amount of 
modelling is always needed and added sophistication is continually being made
to calculations in this approach. There is a clear overlap with the 
Diffraction working group. However, in general the free parameters in a 
dipole model are  determined by a fit to $F_2(x,Q^2)$ data 
and results compared to more exclusive processes. An example of this
was presented by Watt \cite{Wattdip} using an extended dipole model with impact
parameter dependence. This gives a good quality comparison to a wide
variety of exclusive processes, though each of the two different models used 
works better in some cases. The saturation scale for this 
and a number of other approaches, all including heavy flavour contributions, 
is shown in Fig. \ref{fig:satscale} \cite{dipsum}. The saturation scale 
is seen to be at very low $x$ even for $b=0$, falling to lower $x$ as $b$ 
rises (the average for inclusive processes is $b \sim 2-3 \GeV^{-1}$). 
As seen, there are similar results from most other sophisticated and recent 
determinations of parameters using dipole models. 

\begin{figure}[htb]
\vspace{9pt}
  \includegraphics[width=0.45\textwidth]{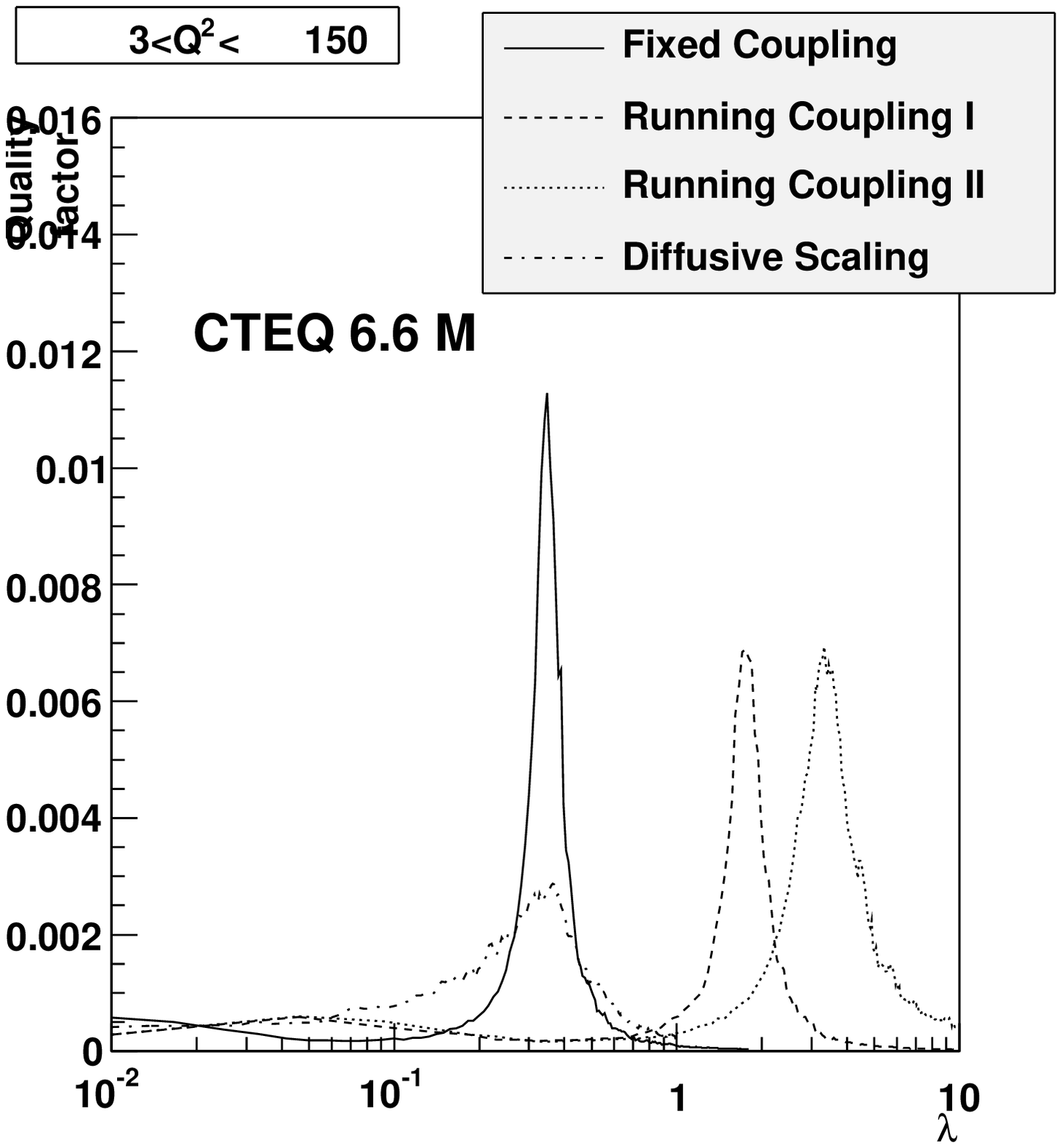}
\caption{The quality factor for various types of geometric scaling using
fixed order QCD and CTEQ PDFs.}
\label{fig:salek}
\end{figure}

The issue of dipole models leads to geometric scaling. It has long been 
emphasised that to a reasonable approximation the HERA inclusive 
cross section can be described in terms of the variable $\tau=Q^2
(x_0/x)^{\lambda}$ \cite{geomscal}, though there are now variations 
of the precise functional form depending on the type of calculation. It
has always been difficult to know how seriously to take this since it 
is broken (by higher orders in calculations, heavy quark contributions, ...) 
by significant amounts. In addition it was shown explicitly that   
geometric scaling can appear from standard DGLAP evolution \cite{Caola}
with sensible boundary conditions. Perhaps more striking was the observation 
that $F_2(x,Q^2)$ 
generated from MRST and CTEQ PDFs display {\it all} types of 
geometric scaling with good quality factors \cite{Beuf}. In this case 
there is no saturation in the generation at all, yet it is displayed by 
the theoretical output, as seen in Fig. \ref{fig:salek}, 
and indeed, more clearly than by real data. This, at the very least,
suggests that 
geometric scaling is a general feature that is generated by saturation,
but also in calculations where saturation is not used, or required.

\section{Perturbative QCD calculations}

There was a great deal of new activity in the field of jet definitions 
at the workshop. It has long been known that the initial cone-based 
jet algorithms are generally infrared unsafe. For example, in seeded cone 
algorithms the appearance of a single very soft particle can 
counterintuitively change the number of jets. In calculations this can 
lead to infrared sensitivity. As shown explicitly by Salam, this has 
quantitative consequences in reality. Formally at some order a divergence 
will appear, i.e. we have a formal series of the form 
\[
\alpha_S^2 + \alpha_S^3 + \alpha_S^4 \times \infty, \nonumber
\]
which in reality with some finite low scale cut-off $\Lambda$, becomes
\[
\alpha_S^2 + \alpha_S^3 + \alpha_S^4 \times \ln(p_t^2/\Lambda^2), \nonumber
\]
and the incorrect last term is of size ${\cal O}(\alpha_S^3)$, so the 
result is meaningless beyond the lowest order. With different IR-unsafe
algorithms and different quantities this breakdown of the expansion appears
at different orders, sometimes at leading order. 

\begin{figure}[htb]
\vspace{9pt}
  \includegraphics[width=0.25\textwidth]{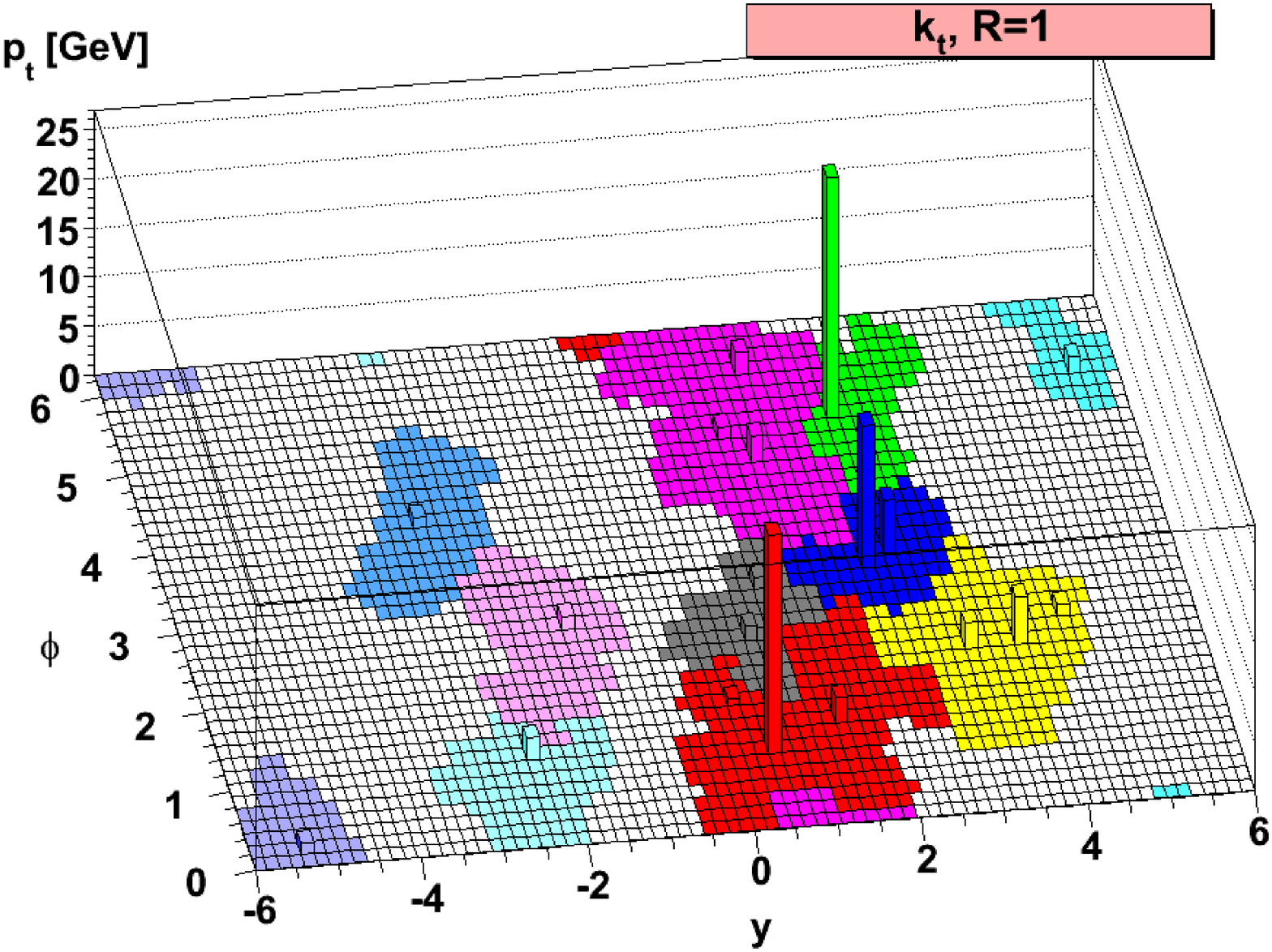}\includegraphics[width=0.25\textwidth]{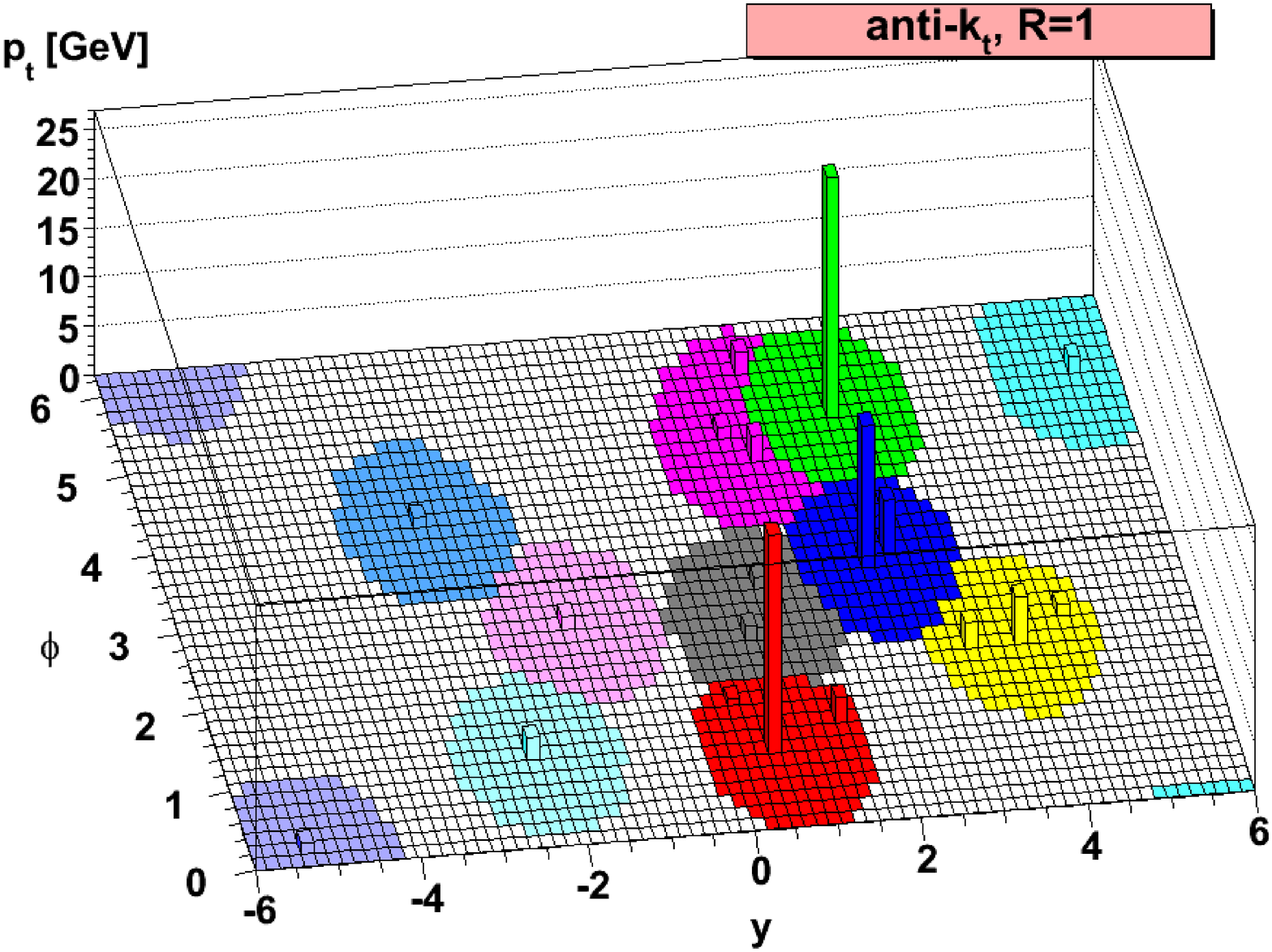}
\caption{Shape of jets using $k_t$ and anti-$k_t$ algorithms.}
\label{fig:antikt}
\end{figure}

A new seedless infrared safe cone algorithm, SISCone \cite{Siscone}
was presented. There was also the introduction of a ``anti-$k_t$ 
algorithm'' \cite{antikt} which is a recombination algorithm based on the
separation
\[
d_{ij} = \min(k_{t,i}^{-2},k_{t,j}^{-2})(\Delta \phi_{ij}^2
+ \Delta \eta_{ij}^2). \nonumber
\]
This has all the useful calculational properties of the IR-safe $k_t$ 
algorithm, but produces much more regularly shaped jets 
by combining all soft partons within a ``cone'' with a hard parton to produce
a cone-like jet definition, as seen in Fig. \ref{fig:antikt}. This is 
useful when using jet area to subtract underlying event and pile up,
where $p_T/A$ is fairly constant except for hard jets \cite{jetarea}.

\begin{figure}[htb]
\vspace{9pt}
  \includegraphics[width=0.45\textwidth]{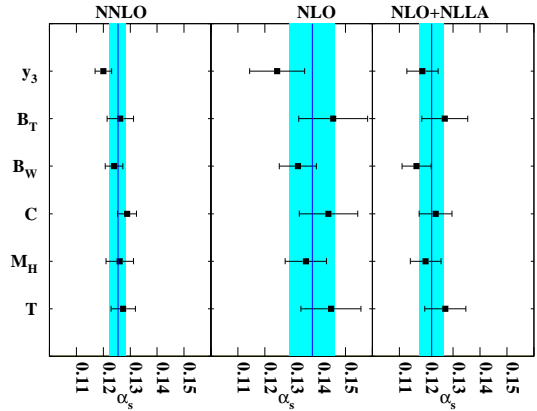}
\caption{Determination of $\alpha_S(M_Z^2)$ from jet shapes.}
\label{fig:ascat}
\end{figure}

Many new developments in the area of calculations 
in perturbative QCD were presented. These included automation of 
NLO calculations \cite{QCD1}, improved parton showering which could be
applied beyond LO \cite{QCD2}, duality between one-loop and single-cut 
phase space integrals \cite{QCD3}, automated one-loop $N$-gluon amplitudes via
unitarity \cite{QCD4}, automated implementation of dipole subtraction
\cite{QCD5} and efficient ways of inputing NLO calculations into analyses
\cite{QCD6,fastnlo}. There was also progress reported at NNLO. 
A full NNLO calculation of jet event shapes in $e^+e^-$ annihilation was 
reported \cite{Gehrmann:2008kh} as was the application to extracting 
$\alpha_S(M_Z^2)$ \cite{Dissertori:2008cn}.  
The results can be seen in Fig. \ref{fig:ascat}. 
The NNLO calculation reduces the uncertainty in the extraction of 
$\alpha_S(M_Z^2)$ compared 
to NLO, but leads to a high value. Combination with additional resummations 
is necessary and nearing completion.  
There was even major progress reported on the full NNLO calculation of 
heavy quark production in hadron-hadron collisions \cite{topnnlo}. 
The exact contributions for virtual corrections to quark annihilations
have been calculated, and it was suggested that  
the total quark contributions may soon be known.

\section{Heavy flavour}

One of the main features of the Heavy Flavours working group
was a detailed comparison of variable flavour number schemes. 
The brief conclusion is that the MSTW and ACOT definitions of 
a General Mass Variable Flavour Number Scheme are very similar,
but there are some outstanding differences in detail which can be thought 
of as higher order effects. A detailed summary can be found in  
\cite{Thorne:2008xf}. It was highlighted that when comparing with the 
precise new data up-to-date definitions and PDF sets should be used. There was 
some discussion on the present unavailability of fixed flavour number scheme 
calculations at NNLO, and preliminary steps in this direction were
presented \cite{Bierenbaum:2008tm}.   

\begin{figure}[htb]
\vspace{9pt}
  \includegraphics[width=0.45\textwidth]{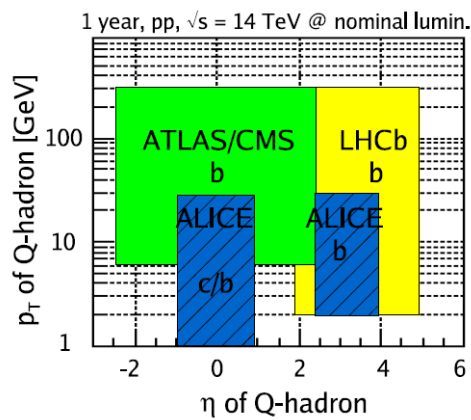}
\caption{Acceptance for heavy flavours at the LHC.}
\label{fig:hfaccept}
\end{figure}

There will be excellent coverage of heavy flavour at 
the LHC, as seen in Fig \ref{fig:hfaccept} and described in 
\cite{HFaccept}, with LHCb and particularly Alice having triggers to 
extremely low $p_T$. The enormous cross section for heavy flavour 
production at the LHC means  
it doesn't really matter that the bandwidth for the trigger for B 
physics is only about $5 \%$. All detectors have a wide ranging heavy flavour 
physics programme. Measurements of open heavy flavour production, 
heavy flavour jets, quarkonium production, oscillations of $B$ mesons 
and rare decays will all be covered by at least one of the experiments,
and usually more.  
Measurements of heavy flavours down to low $p_T$ and at higher 
rapidity will test QCD in the same manner as the low-mass Drell-Yan
production. This will constrain small-$x$ PDFs, check for small-$x$ 
resummations \cite{Ball:2001pq}, saturation {\it etc.}
However, theory is even more uncertain in this case. Predictions
with largely similar underlying bases show uncertainties of factors of 
up to 2 \cite{HFaccept}. Using more extreme theoretical approaches 
there is more scope for variations, 
so there is a lot of work for theorists here. 

It was particularly noted that there is the possibility of very quick 
results on heavy meson production (Lytken). 
With $10$ pb$^{-1}$ it will already be possible
to measure $J/\psi$ polarisation to the same precision as the Tevatron
with 1.3 fb$^{-1}$ but reaching to higher $p_T$. The same precision 
for $\Upsilon$ polarisation studies can be reached after $100$ pb$^{-1}$
and should help clarify the situation on the appropriate theoretical 
framework, with no clear-cut result being available using present data 
\cite{:2008za}. 

\section{Diffraction}

\begin{figure}[htb]
\vspace{9pt}
  \includegraphics[width=0.48\textwidth]{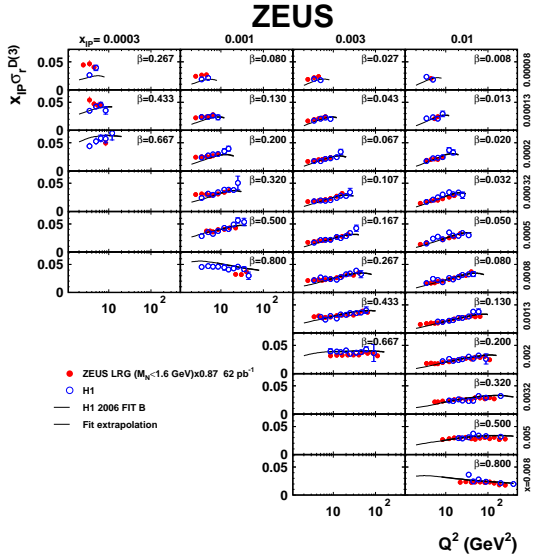}
\caption{The large rapidity gap diffractive data from H1 and ZEUS.}
\label{fig:diffdata}
\end{figure}

There have recently been major 
improvements in consistency between measurements,
e.g. the comparison of the H1 \cite{Aktas:2006hy} 
and ZEUS \cite{Ruspa:2008qj} inclusive diffractive cross sections 
using the 
Large Rapidity Gap definition in Fig. \ref{fig:diffdata}. 
In addition, inclusion of jet production has stabilised the results of 
the H1 fits 
dramatically \cite{Aktas:2007bv}, 
in fact bringing them near to the result in the alternative 
MRW approach \cite{Martin:2006td}. 
The level of agreement between experiments is such that it would now 
be appropriate to produce averaged data, and 
(subsequently) combined fits.
The results of these could then be used at the LHC.

\begin{figure}[htb]
\vspace{9pt}
  \includegraphics[width=0.48\textwidth]{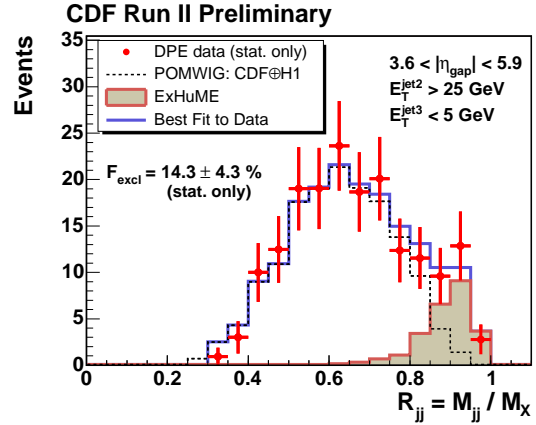}
\caption{Dijet production at CDF.}
\label{fig:exhume}
\end{figure}

However, while diffractive PDFs obey factorization, i.e. 
one can determine evolution and combine with hard coefficient 
functions, the 
factorization is not universal. 
Very simple application of extracted PDFs to Tevatron data
does not, and was never expected to work, giving results $\sim 10$ times too 
high \cite{Affolder:2000vb}. 
Factorization is known to be broken in hadronic diffraction due to 
soft interaction filling in gaps in both initial and final states. 
This can be interpreted as a {\it phenomenological} 
``gap survival'' probability. 
It is hoped this approach can give some reasonable 
accuracy for prediction of LHC processes, though the survival 
probabilities for different processes are   
not all alike. There is dependence on the nature of 
the basic process, kinematical configuration, cuts {\it etc.} 
\cite{Khoze:2001xm}. 
Predictions for diffractive processes at LHC can be tested 
with measurements at Tevatron \cite{Aaltonen:2007hs}, and evidence 
for doubly diffractive dijet production is seen in Fig. \ref{fig:exhume}. 
The excess in dijet with a very large fraction of the total invariant mass
of the final state is well described by the generator 
ExHuMe \cite{Monk:2005ji} based on the KMR calculations \cite{Khoze:2001xm} 
with a $4.5\%$ gap survival probability.

In principle there is a good test of factorisation 
also in diffractive photoproduction. 
The naive guess is that the direct contribution (like DIS) satisfies 
factorisation, but the resolved (like hadronic processes) 
has a gap survival $\sim 0.3$. 
Initially ZEUS \cite{:2007yw} and H1 data \cite{Aktas:2007hn} 
did not agree well, and
there was a suggestion of a suppression $\sim 0.5$ at all $x_{\gamma}$
in the latter but not the former. 
There have been recent improvements in understanding data differences
with a suggestion of $E_T$ dependence, see e.g. \cite{Schoning:2008yu}.
It appears that suppression with a gap survival factor 
of $\sim 0.4$ independent of $x_{\gamma}$ works best, but this is 
still a matter of investigation \cite{Klasen:2008ah}.

\section{Monte Carlo, tools}

There have been a considerable number of improvements in 
Monte Carlo generators 
and associated tools during the time of the Workshop series, and many updates 
were noted. There were various developments in parton showers, e.g. 
\cite{QCD2} and \cite{Jadach:2008nu}. In particular a report on
two new dipole showers {\tt ADICIC++} \cite{Winter:2007ye} and 
{\tt CSSHOWER++} \cite{Schumann:2007mg} used in {\tt SHERPA} was presented. 

\begin{figure}[htb]
\vspace{9pt}
  \includegraphics[width=0.48\textwidth]{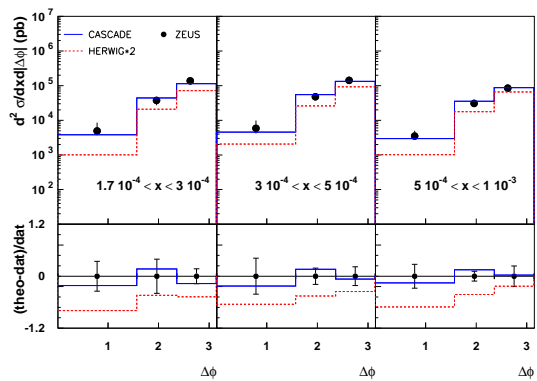}
\caption{Angular correlations for 3-jet production comparing {\tt CASCADE} and
{\tt HERWIG++}.}
\label{fig:jetcorr}
\end{figure}

There was a comprehensive update report on the 
status of all the major Monte Carlo generators and 
associated tools. I begin with {\tt CASCADE} \cite{Jung:2001hx}.
This is very different from standard Monte Carlo generators, and is based on 
generation of unintegrated PDFs via the CCFM equation. As such it has 
advantages, such as better treatment of high-energy logarithms, and
is likely to be useful for processes sensitive to parton $k_T$. 
Successes for angular correlations in jets were outlined 
\cite{Hautmann:2008vd}, and an example can be seen in Fig. \ref{fig:jetcorr}.
However, there are corresponding disadvantages, e.g. it is less complete in 
the high-$x$ regime, in particular currently not fully including quark 
contributions. Additionally knowledge of the unintegrated PDFs is 
required. All these may be important, and the most appropriate regime for 
the use of {\tt CASCADE} should always be borne in mind, though
improvements are continuing \cite{Deak:2008iz}.

\begin{figure}[htb]
\vspace{9pt}
  \includegraphics[width=0.48\textwidth]{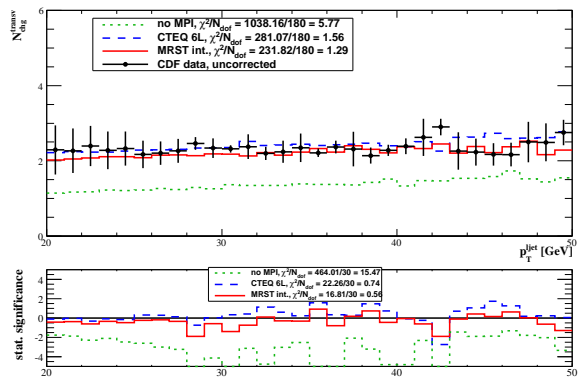}
\caption{Predictions for number of charged particles using  the 
multiparticle interaction model in {\tt HERWIG++}.}
\label{fig:mpiherwig}
\end{figure}

The {\tt ARIADNE} generator \cite{Lonnblad:1992tz} 
is based on a colour dipole cascade model. This has been completely
rewritten in {\tt C++} and has been validated for $e^+e^-$ annihilation, 
but some more work is needed for use at the LHC. As well as new parton 
showers,
{\tt SHERPA} \cite{Gleisberg:2008ta} includes new cluster fragmentation 
\cite{Winter:2003tt} and a 
new method for calculating high multiplicity 
cross sections {\tt COMIX} \cite{Gleisberg:2008fv}. Significant
developments in the new update of {HERWIG++} \cite{Bahr:2008pv} were 
reported, e.g. new models of decays, inclusion of more beyond the standard 
model physics (MSSM and extra dimension models), and inclusion of new
hard processes.  In particular a new model of multiple partonic interactions
was presented \cite{Bahr:2008dy}, 
and a distinct improvement in the description of 
the underlying event at the Tevatron is seen, illustrated in 
Fig. \ref{fig:mpiherwig}. A summary of {\tt PYTHIA8} \cite{Sjostrand:2007gs} 
was reported, emphasising that it is a completely new generator. The 
improved features include interleaved $p_T$-ordered evolution, more 
underlying event processes, updated decay chains, the possibility for two
selected hard interactions in the same event and the possibility to use 
one PDF sets for the hard process and another for the rest.

\begin{figure}[htb]
\vspace{9pt}
  \includegraphics[width=0.48\textwidth]{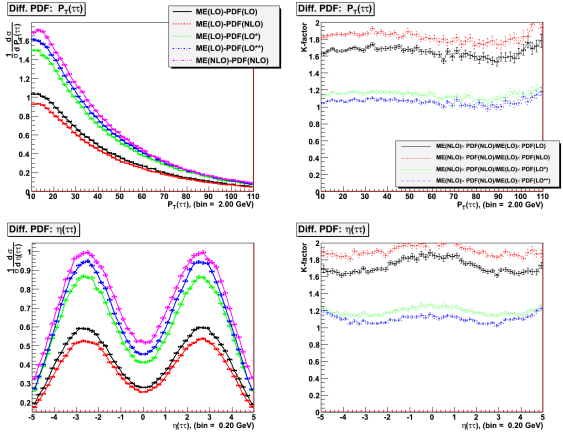}
\caption{Predictions for $H \to \tau^+\tau^-$ at full NLO and at LO with 
various choices of PDF.}
\label{fig:higgsratio}
\end{figure}

The last point is due to the suggestion that some PDFs, e.g. LO,
are appropriate for some processes, e.g. underlying event, and others, e.g.
NLO for others, e.g. $W,Z$ production \cite{Campbell:2006wx}. 
This brings me in a full circle back to PDFs. One of the issues discussed in 
detail in the series of workshops and in the offshoot PDF4LHC meetings has 
been the question of the PDFs to be used in LO Monte Carlo generators.
The problems in using either standard 
LO or NLO PDFs in all cases was highlighted in \cite{Sherstnev:2007nd}. 
One alternative approach suggested by Jung is to obtain PDFs by 
fitting directly using Monte Carlo generators. Work on this project (PDF4MC) 
has begun using {\tt RAPGAP} \cite{Jung:1993gf} to fit to HERA data 
on $F_2$, charm production and dijets. This is self-consistent, but 
will take a lot of time 
to do thoroughly and with anything like as wide constraints as the global 
fits, and a good simultaneous fit to all 
data will be difficult to obtain. 
Indeed, the main problem with working entirely at 
LO is that most NLO corrections are positive, many large, rendering
globally good fits and predictions impossible. This has led to 
the production of modified LO partons for LO 
Monte Carlo generators \cite{Sherstnev:2007nd} where LO* PDFs are enhanced
by allowing momentum violation and 
use of the NLO strong coupling. A further extension, LO**, has 
a Monte Carlo inspired choice of scale in the coupling 
\cite{Sherstnev:2008dm}. As seen in Fig. \ref{fig:higgsratio}   
these modified PDFs often lead to the best match to full 
NLO predictions. More generally, they provide results which are consistently 
very close to the full NLO result, so should be useful if a LO calculation is 
used due to time constraints or lack of a NLO generator. 
Similar modified LO sets are in preparation by CTEQ 
(presented at PDF4LHC in June 2008).  
These are also based (sometimes) on momentum violation in the input 
PDFs and sometimes in this case fitting to LHC pseudodata, possibly using 
$K$-factors for normalisations. This project of PDFs for use in Monte Carlos
is one of many related to this series of workshops which is ongoing. 

\section{Conclusions}

The objectives of the series of workshops, as defined by the organisers, were:

\begin{itemize}

\item To encourage and stimulate transfer of knowledge between the HERA and
LHC communities and establish an ongoing interaction.

\item To increase the quantitative understanding of the implications of
HERA measurements for LHC physics. 

\item To encourage and stimulate theory and phenomenological efforts. 

\item  To examine and improve theoretical and experimental tools. 

\item To identify and prioritise those measurements to be made at HERA which 
have an impact on the physics reach of the LHC. 

\end{itemize}

The Workshop has clearly been very successful in all of these, and 
the organisers 
deserve congratulations and thanks. However, I have a couple of 
minor comments to add. 
As the summary shows, despite the name of the Workshop, Tevatron results 
and people have also made a big contribution and are 
involved in collaborations. 
Secondly, it was  noticeable that quite a few people at the 
LHC who looked in at the first meeting 
decided to forget about importance of QCD. It is likely they will 
be obliged to remember when  data starts to appear.

It is now the end of the HERA-LHC Workshop series in this form. 
However, the work started will undoubtedly carry on. 
The Parton Density Functions working group has
developed into a PDF4LHC committee, and series of smaller workshops,
which took place in February, July and September 2008, and will continue.   
There has already been a MC4LHC workshop, and there is continuing effort in 
the MCNET network. It would seem as though there is definitely scope for 
something similar along the lines of jets4LHC, diffraction4LHC, candb4LHC
{\it etc.}. We really should build on the good work started and many 
collaborations established by the HERA-LHC Workshop series.

\section*{Acknowledgements}

I would like to thank Albert de Roeck and Hannes Jung for organising the
HERA-LHC Workshop series so effectively, and the organisers of the Ringberg
Workshop for inviting me to give this summary.

\end{document}